\documentclass[useAMS,usenatbib]{mnras}
\usepackage{Preamble}
\usepackage{graphicx,amssymb,amsmath,times,verbatim,pdfpages,multirow,booktabs,array,dcolumn}
\usepackage[flushleft]{threeparttable}
\usepackage[countmax]{subfloat}
\usepackage{xcolor}
\usepackage{ulem}
\usepackage{soul}
\usepackage{lineno}

\newcommand{\blue}[1]{\textcolor{blue}{#1}}

\title[Optical variability of TXS 0506+056]{Multi-band optical variability on diverse timescales of the TeV blazar TXS 0506+056, the first cosmic neutrino source}
\author[Dhiman et al.]{Vinit Dhiman$^{1,2}$\thanks{Email: dhiman@aries.res.in}\orcidlink{0000-0002-8105-4566}, 
Alok C. Gupta$^{3,1}$\thanks{Email: acgupta30@gmail.com}\orcidlink{0000-0002-9331-4388}, 
Rumen Bachev$^{4}$\orcidlink{0000-0002-0766-864X}, 
Paul J. Wiita$^{5}$\orcidlink{0000-0002-1029-3746},  
Sergio A. Cellone$^{6,7}$\orcidlink{0000-0002-3866-2726}, 
\newauthor
A. Strigachev$^{4}$, 
Haritma Gaur$^{1}$\orcidlink{0000-0002-6629-8490}, 
A. Darriba$^{8,9}$, D. P. Bisen$^{2}$, G. Locatelli$^{10}$, L. A. Mammana$^{6,7}$, 
E. Semkov$^{4}$\orcidlink{0000-0002-1839-3936} \\
\\
$^{1}$Aryabhatta Research Institute of Observational Sciences (ARIES), Manora Peak, Nainital 263001, India \\
$^{2}$School of Studies in Physics \& Astrophysics, Pt.\ Ravishankar Shukla University, Amanaka G.E. Road, Raipur 492010, India \\
$^{3}$Key Laboratory for Research in Galaxies and Cosmology, Shanghai Astronomical Observatory, Chinese Academy of Sciences, Shanghai 200030, China \\
$^{4}$Institute of Astronomy and National Astronomical Observatory, Bulgarian Academy of Sciences, 72 Tsarigradsko Shosse Blvd., 1784 Sofia, Bulgaria \\
$^{5}$Department of Physics, The College of New Jersey, 2000 Pennington Rd., Ewing, NJ 08628-0718, USA \\
$^{6}$Complejo Astronómico El Leoncito (CASLEO), CONICET-UNLP-UNC-UNSJ, Av. España 1512 (Sur), J5402DSP, San Juan, Argentina \\
$^{7}$Facultad de Ciencias Astronómicas y Geofísicas, Universidad Nacional de La Plata, Paseo del Bosque, B1900FWA, La Plata, Argentina \\
$^{8}$American Association of Variable Star Observers (AAVSO), 49 Bay State Road, Cambridge, MA 02138, USA \\
$^{9}$Group M1, Centro Astronómico de Avila, Madrid, Spain \\
$^{10}$Maritime Alps Observatory, Cuneo, Italy}

\begin{document}
\label{firstpage}
\maketitle
\begin{abstract}
\noindent
We report the first extensive optical flux and spectral variability study of the TeV blazar TXS\,0506$+$056 on intra-night to long-term  timescales using {\it BVRI} data collected over 220 nights between January 21, 2017 to April 9, 2022 using 8 optical ground-based telescopes. In our search for intraday variability (IDV), we have employed two statistical analysis techniques,  the nested ANOVA test and the power enhanced $F$-test. We found the source was variable in 8 nights out of 35 in the $R$-band and in 2  of 14 in the $V$-band yielding Duty Cycles (DC) of 22.8\% and 14.3\%, respectively. Clear colour variation in {\it V -- R} was seen in only 1 out of 14 observing nights, but no IDV was found in the more limited {\it B, I}, and {\it B -- I} data. During our monitoring period the source showed a 1.18 mag variation in the {\it R}-band and similar variations are clearly seen at all optical wavelengths. We extracted the optical ({\it BVRI}) SEDs of the blazar for 44 nights  when observations were carried out in all four of those wavebands. The mean spectral index ($\alpha$) was determined to be $0.897 \pm 0.171$. 
\end{abstract}


\section{Introduction} \label{sec:introdues}
\noindent
Blazars are Active Galactic Nuclei (AGN) characterized by  highly collimated relativistic jets with half opening angles $\lesssim$ 5$^{\circ}$ closely aligned with the observer’s line of sight \citep[e.g.,][]{1995PASP..107..803U}. The Doppler enhanced intense non-thermal radiation from this jet dominates the spectral energy distribution (SED) from radio to very high energy (VHE) $\gamma$-ray energies. Blazars are usually considered to be comprised of BL Lacertae (BL Lac) objects which show featureless spectra or very weak emission lines (equivalent width EW $\leq  5\,$\AA) \citep{1991ApJ...374...72S,1996MNRAS.281..425M}, and flat spectrum radio quasars (FSRQs) with prominent emission lines in their composite optical/UV spectra \citep{1978PhyS...17..265B,1997A&A...327...61G}. Blazars show flux and spectral variability across the entire electromagnetic (EM) spectrum, emit predominantly non-thermal radiation showing strong polarization ($>3\%$) from radio to optical frequencies, and usually have core dominated radio structures. \\
\\
The multi-wavelength (MW) SEDs of blazars in log($\nu F_{\nu}$) vs. log($\nu$) representation show double-humped structures in which the low energy hump peaks in infrared (IR) through X-ray bands while the  high energy hump peaks at $\gamma$-ray energies \citep{1998MNRAS.299..433F}. The location of SED peaks are often used to classify blazars into two sub-classes, namely LBLs (low-energy-peaked blazars) and HBLs (high-energy-peaked blazars). In LBLs, the first hump peaks in IR to optical bands and the second in GeV $\gamma$-ray energies \citep{1995MNRAS.277.1477P}. Whereas in HBLs, the first hump peaks in UV to X-ray bands and the second in up to TeV $\gamma$-ray energies  \citep{1995MNRAS.277.1477P}. The emission of the lower energy SED hump is due to synchrotron radiation which originates from relativistic electrons in the jet. The high-energy hump can be attributed to two mechanisms. One of these is inverse Compton (IC) scattering of low-energy photons by the same electrons responsible for the synchrotron emission (synchrotron-self Compton, SSC) or external photons (external Compton, EC), collectively known as the leptonic model \citep[e.g.,][]{2007Ap&SS.307...69B}. The other mechanism is emission from relativistic protons or muon synchrotron radiation, referred to as the hadronic model \citep[e.g.,][]{2003APh....18..593M}. \\  
\\
Flux variability over a wide range of timescales is one of the definitional properties of blazars. On the basis of the timescales over which it is observed, blazar variability can be divided into three somewhat arbitrary categories: microvariability \citep{1989Natur.337..627M} or intra-day variability (IDV) \citep{1995ARA&A..33..163W} or intra-night variability  \citep{1993MNRAS.262..963G} (occurring on a timescale of a few minutes to less than a day); short-term variability (STV; taking place on a timescale of days to months); and long-term variability (LTV; over a timescale of several months to years or even decades) \citep{2004A&A...422..505G}.\\

\begin{table*}
%\centering
\caption{Details of telescopes and instruments used.} 
\scalebox{1.0}{
\label{tab:tel_log}             
\begin{tabular}{l c c c c}          
\hline \hline                		
                     & A1           	     & A2		           & B     		         & S	   \\\hline 
Telescope            & 1.30\,m DFOT       & 1.04\,m ST          & 60\,cm AO	                   & 1.\,3m Modified RC \\\hline 
CCD Model            & Andor 2K CCD       & STA4150		        & FLI PL9000	               & Andor DZ936 BXDD \\ 
Scale (arcsec/pixel) & 0.535  	          & 0.264	            & 1.0		                   & 0.2829	  \\    
Field (arcmin$^2$)   & $18\times18$ 	  & $16\times16$	    &$16.8\times16.8$	           & $9.6\times9.6$	\\	 
\hline   
                     &    R2	             & C                   &  Sp                &    I    \\\hline
Telescope            & 2\,m RC NAO        & 0.6\,m HSH   &  35.6 cm Schmidt Cassegrain  & 25cm Schmidt Cassegrain \\\hline
CCD Model            & VersArray:1300B    & SBIG STL-1001E      &  ATIK 383L+ Monochrome       & QHY9 (KAF8300) monochrome \\
Scale (arcsec/pixel) & 0.258			  & 0.51                &  1.38                        & 0.782 \\
Field (arcmin$^2$)   & $5.76\times5.76$   & $9.3\times9.3$      &  $25.46\times19.16$          & $43.7\times33.1$\\
\hline
\end{tabular}
}\\
\footnotesize

A1: 1.3\,m Devasthal Fast Optical Telescope (DFOT) at ARIES, Nainital, India\\
A2: 1.04\,m Samprnanand Telescope (ST), ARIES, Nainital, India\\
B: 60\,cm Cassegrain Telescope at Astronomical Observatory Belogradchik, Bulgaria \\
S: 1.3\,m Skinakas Observatory, Crete, Greece \\
R2: 2\,m Ritchey-Chretien telescope at National Astronomical Observatory Rozhen, Bulgaria \\
C : 0.6\,m HSH classic Cassegrain at CASLEO, Argentina \\
Sp: 35.6 cm Telescope at Observatorio Astronomico Las Casqueras, Spain \\
I: 25 cm Telescope at Maritime Alps Observatory Cuneo, Italy \\
\end{table*}

\noindent
Colour-magnitude diagrams (CMDs) for blazars can be analyzed to find any colour trends accompanying brightness changes. Three types of CMD behaviour could be discerned: redder-when-brighter (RWB), bluer-when-brighter (BWB), and achromatic. FSRQs mostly show RWB chromatism because the contribution of the accretion disc to the total emission is significant \citep{2006A&A...450...39G,2012MNRAS.425.3002G}. The BWB behaviour seen in many BL Lacs is thought to arise from processes occurring in the relativistic jet, such as particle acceleration and cooling in the framework of the shock-in-jet model \citep[e.g.,][]{1985ApJ...298..114M,1998A&A...333..452K}. Alternatively, the BWB chromatism could arise from a Doppler factor variation on a convex spectrum \citep[e.g.,][]{2004A&A...421..103V,2007A&A...470..857P}. Finally, achromatic behaviour is frequently interpreted as being due to the variations of the Doppler factor, which are most likely explained in a framework involving changes in the viewing angle to the dominant emission region \citep[e.g.,][]{2002A&A...390..407V,2006A&A...450...39G,2012MNRAS.425.3002G,2015MNRAS.450..541A,2016MNRAS.455..680A}.\\
\\
TXS\,0506$+$056 is registered as a blazar in the Texas Survey of Radio Sources catalog \citep{1996AJ....111.1945D}. The first detection of a high-energy neutrino event from a blazar was reported from TXS\,0506$+$056 on 22 September 2017 by the IceCube collaboration and was coincident in  direction and time with a $\gamma-$ray flare \citep{2018Sci...361..147I}. Prompted by this discovery, an investigation was carried out of 9.5 years of IceCube neutrino observations to search for excess emission at the position of the blazar, and an excess of high-energy neutrino events between September 2014 and March 2015 at energies around 290 TeV at a $3.5\,\sigma$ level was indeed detected \citep{2018Sci...361.1378I}. This object is the highest energy $\gamma$-ray-emitting blazar among those detected by the Energetic Gamma Ray Experiment Telescope (EGRET) satellite in the $\gamma$-ray energy range (30 MeV$-$30 GeV)  \citep{2001AIPC..587..251D}. It is identified as a jet-dominated in the low-hard state during neutrino flaring in 2014/2015, and so provides evidence for the blazar jet acting as an accelerator of cosmic-ray particles which produce neutrinos \citep{2018MNRAS.480..192P}. TXS\,0506$+$056 was independently detected at high energy $\gamma$-rays with the Large Area Telescope (LAT) onboard the Fermi satellite \citep{2017ATel10791....1T}, the MAGIC telescope \citep{2018ApJ...863L..10A}, and the AGILE $\gamma$-ray telescope \citep{2019ApJ...870..136L}, which strengthens the case for  TXS\,0506$+$056 being a very high energy (VHE) $\gamma$-rays emitting BL\,Lac as well as a neutrino emitting source. However, \citet{2019MNRAS.484L.104P} claimed that TXS\,0506$+$056 is a masquerading BL\,Lac, i.e., a FSRQ with hidden broad lines and a standard accretion disc that is outshined by the jet emission. During the intensive follow-up observations, the redshift of TXS\,0506$+$056 was successfully determined to be $z = 0.3365$ \citep{2018ApJ...854L..32P}. The analysis of single-dish 15\,GHz radio flux densities from the Owens Valley Radio Observatory (OVRO) spanning between 2008 and 2018 indicates that the core of TXS\,0506$+$056 is in a highly flaring state coincident with the neutrino event 170922A \citep{2019A&A...630A.103B, 2019MNRAS.483L..42K}. In this paper we are  reporting an extensive optical variability study of the first neutrino emitting TeV blazar TXS\,0506$+$056 on diverse timescales.\\
\\
This paper is organised as follows: Section 2 provides an overview of the telescopes, photometric observations, and the data reduction procedure. Analysis techniques we used to search for flux variability and correlations between bands are discussed in Section 3. Results of our study are reported in Section 4. A discussion and conclusions  are provided in Section 5. 

\section{Observations and Data Reduction} \label{sec:data}
\noindent
Optical photometric observations of the TeV blazar TXS\,0506$+$056 were carried out using 8  ground-based telescopes. Two telescopes are located in India: the 1.04\,m Sampurnanand Telescope (ST), and the 1.3\,m Devasthal Fast Optical Telescope (DFOT), ARIES, Nainital. Both of these telescopes are equipped with CCD detectors and broadband Johnson {\it UBV} and Cousins {\it RI} filters. The source was observed with alternate observations in the $V$ and $R$ bands for a total of 37 nights between November 7, 2019 and January 31, 2021. One or two $B$ and $I$ image frames were also taken on each night of observations.\\
\\
Observations of this source with the 60\,cm Cassegrain telescope located at the Astronomical Observatory Belogradchik, Bulgaria, were carried out over the course of 40 nights from \ 10 October 2018 to 17 August 2020, consisting on a single optical frame in $B$, $V$, $R$, and $I$ bands each night. These 40 nights of observations were presented in \citet{2021BlgAJ..34...79B}. The 2-m Ritchey-Chretien telescope at the National Astronomical Observatory Rozhen, Bulgaria, observed only a single night in $V$, $R$, and $I$ bands on 17 August 2020. Observations of the source with the 1.3\,m modified RC telescope at Skinakas Observatory, Crete, Greece were taken during six nights, 26 -- 31 August  2019 in the optical {\it B, V, R,} and {\it I} bands.  \\
\\
Three additional nights of observations were taken with the 0.6\,m Helen Sawyer Hogg (HSH) telescope at CASLEO, Argentina (on loan from the University of Toronto, Canada) in {\it B, V, R,} and {\it I} bands. Additional {\it V}-band observations from AAVSO\footnote{\url{https://app.aavso.org/vsp/}} (American Association of Variable Star Observers) were carried out from amateur astronomers' two telescopes in Spain and Italy. \\
\\
The technical parameters and instrumental details are summarized in Table \ref{tab:tel_log}. A total of 220 nights of optical photometric observations of TXS\,0506$+$056 were carried out between 21 January 2017 and 9 April  2022. The AAVSO data are included,  with most of these observations being done in the {\it V} and {\it R} bands. In 35 nights, the observation duration is $\geq$ 1h, which we use to study  IDV behaviour of the blazar. {\it R} band observations were carried out in each of them but they were performed in the {\it B, V,} and {\it I} bands in only some of the nights. Considering these 35 nights of observations, we obtained data for 35, 14, 7, and 6 nights, respectively, to look for  IDV in the {\it R, V, I}, and {\it B} bands. During 44 nights we have at least 1 frame in all four optical bands, which are useful for studying this portion of the spectral energy distribution (SED). The observation log is provided in Table \ref{tab:obs_log}. \\
\\
For the preliminary processing of the raw data, we used standard procedures of IRAF\footnote{Image Reduction and Analysis Facility (IRAF) is distributed by the National Optical Astronomy Observatory, which is operated by the Association of Universities for Research in Astronomy (AURA) under a cooperative agreement with the National Science Foundation.} software, following the steps described below. For image pre-processing, we generated a master bias frame for each observing night. This master bias was subtracted from all twilight flat frames and all source image frames taken on that night. A master flat was generated for each passband by taking the median of all the sky flat frames and then normalizing the master flat. Next, to remove pixel-to-pixel inhomogeneity, the source image was divided by the normalized master flat. Finally, cosmic ray removal was carried out for all source image frames using the IRAF task {\sl cosmicrays}. To find the instrumental magnitudes of TXS 0506$+$056 and its comparison stars, we performed the concentric circular aperture photometry technique with the  DAOPHOT\footnote{Dominion Astrophysical Observatory Photometry} II software \citep{1987PASP...99..191S,1992ASPC...25..297S}. For aperture photometry, we explored four different concentric aperture radii defined in terms of the Full Width at Half Maximum (FWHM): 1 $\times$ FWHM, 2 $\times$ FWHM, 3 $\times$ FWHM, and 4 $\times$ FWHM for every night. After examining the results for these different aperture radii, we found that setting the aperture radii = 2 $\times$ FWHM provided the best S/N, so we used this value for our final analysis. \\
\\
Each night we observed more than three local standard stars on the same field. Then we selected three non-varying standard stars (stars A, C and D) from Figure 4 of \citet{2021BlgAJ..34...79B} that are of nearly the same magnitude and colour as the source, to avoid any error occurring from differences in the photon statistics in the differential photometry of the source. Since the magnitudes of TXS 0506$+$056 and the standard stars were obtained simultaneously under the same air mass and weather conditions, there is no need for correction of atmospheric extinction. Finally, a comparison star (Star C) was used to calibrate the instrumental magnitude of the TeV blazar TXS 0506$+$056. Data were collected from different telescopes in the form of instrumental magnitudes of the blazar and reference stars, so that we can apply the same calibration procedures and analysis to all the data sets.

\begin{table*}
\caption{Observation log for TXS\,0506$+$056.} 
\scalebox{1.0}{
\label{tab:obs_log}
\begin{tabular}{l c c c c c}           
\hline \hline
Observatory  & Country   &  Telescope   & Observation Duration     & No. of Nights & Data points \\
             &           & Size (in cm) &                          &               & $B$, \ $V$, \ $R$, \ $I$ \\\hline
ARIES        & India     & 130          & 2019-12-28 to 2020-12-13 &  11           &  9, 154, ~929, 12 \\
ARIES        & India     & 104          & 2019-11-07 to 2021-01-31 &  26           & 11, 160, 1080, 45  \\
Belogradchik$^{*}$ & Bulgaria  &  60          & 2018-10-08 to 2020-04-16 &  40           & 30, ~30, ~~40, 30 \\
Skinakas     & Greece    & 130          & 2021-08-26 to 2021-08-31 &   6           & 60, ~60, ~~60, 60  \\
Rozen        & Bulgaria  & 200          & 2020-08-17 to 2020-08-17 &   1           & ~0, ~~5, ~~~5, ~5  \\
CASLEO       & Argentina &  60          & 2020-11-20 to 2020-11-24 &   3           & 13, ~12, ~113, 12  \\
AAVSO$^{**}$  & Spain     &   35.6           & 2017-01-21 to 2022-04-09 & 105           & ~0, 105, ~~~0, ~0   \\ 
             & Italy     &  25            & 2020-01-17 to 2021-03-01 &  28           & ~0, ~28, ~~~4, ~0  \\\hline
\end{tabular}
}\\
\footnotesize
$^{*}$: The data is also presented in \citet{2021BlgAJ..34...79B}.
$^{**}$: AAVSO - American Association of Variable Star Observers  
\end{table*}

\begin{figure*}
%\ContinuedFloat
\centering
\includegraphics[scale=0.6]{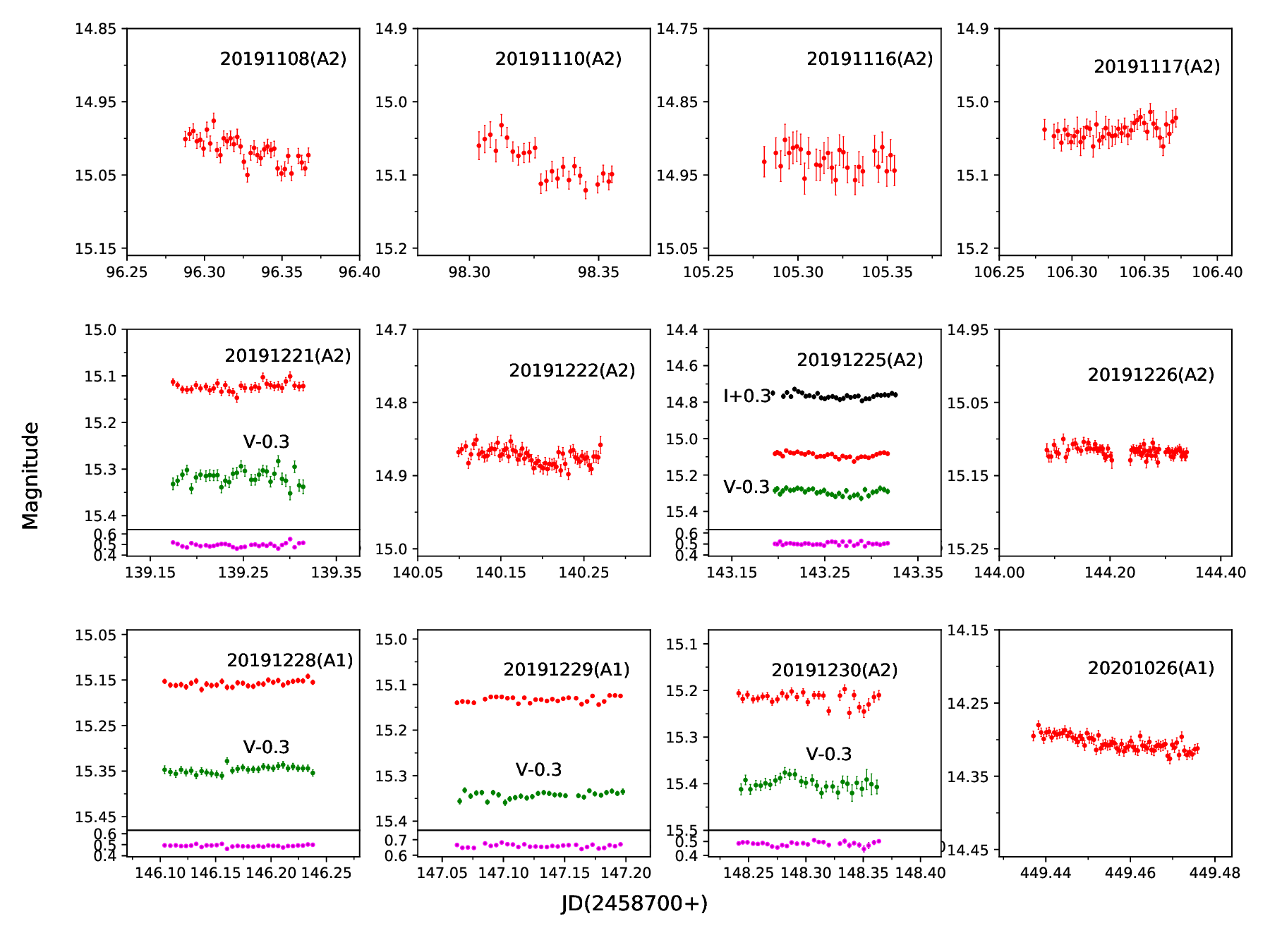}
\vspace*{-0.4in}
\caption{Nightly optical  light curves of the TeV blazar TXS 0506$+$056, $B$-band LCs in blue, $V$-band LCs in green, $R$-band LCs in red, $I$-band LCs in black, $V-R$ colour LCs in magenta, labeled with its observation date and telescope code (from Table 1).}
\label{fig:idv1}
\end{figure*}

\begin{figure*}
\ContinuedFloat
\centering
\includegraphics[scale=0.6]{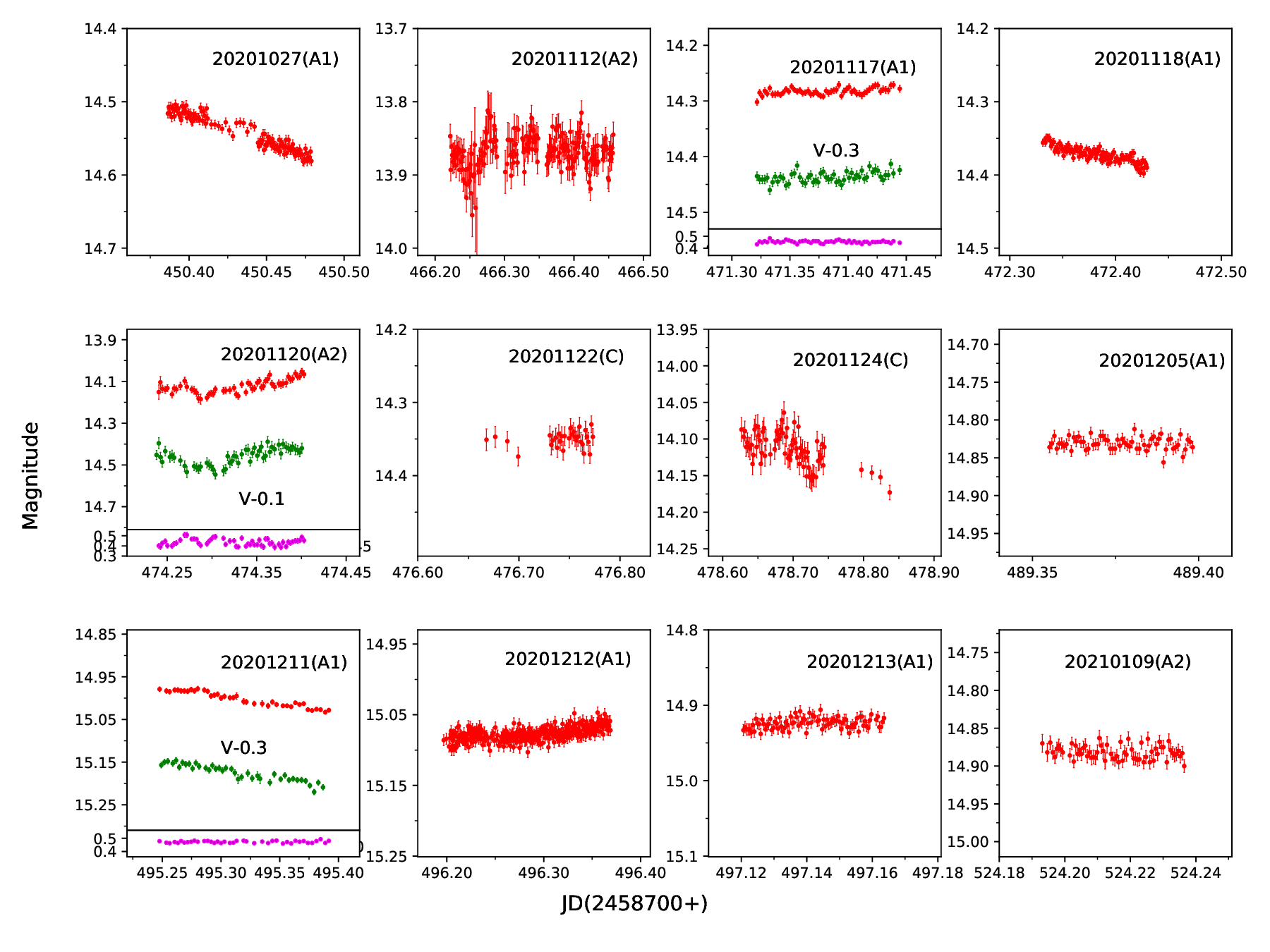}
\vspace*{-0.4in}
\caption{Continued} 
%\label{fig:idv2}
\end{figure*}

\begin{figure*}
\ContinuedFloat
\centering
\vspace*{-0.4in}
\includegraphics[scale=0.6]{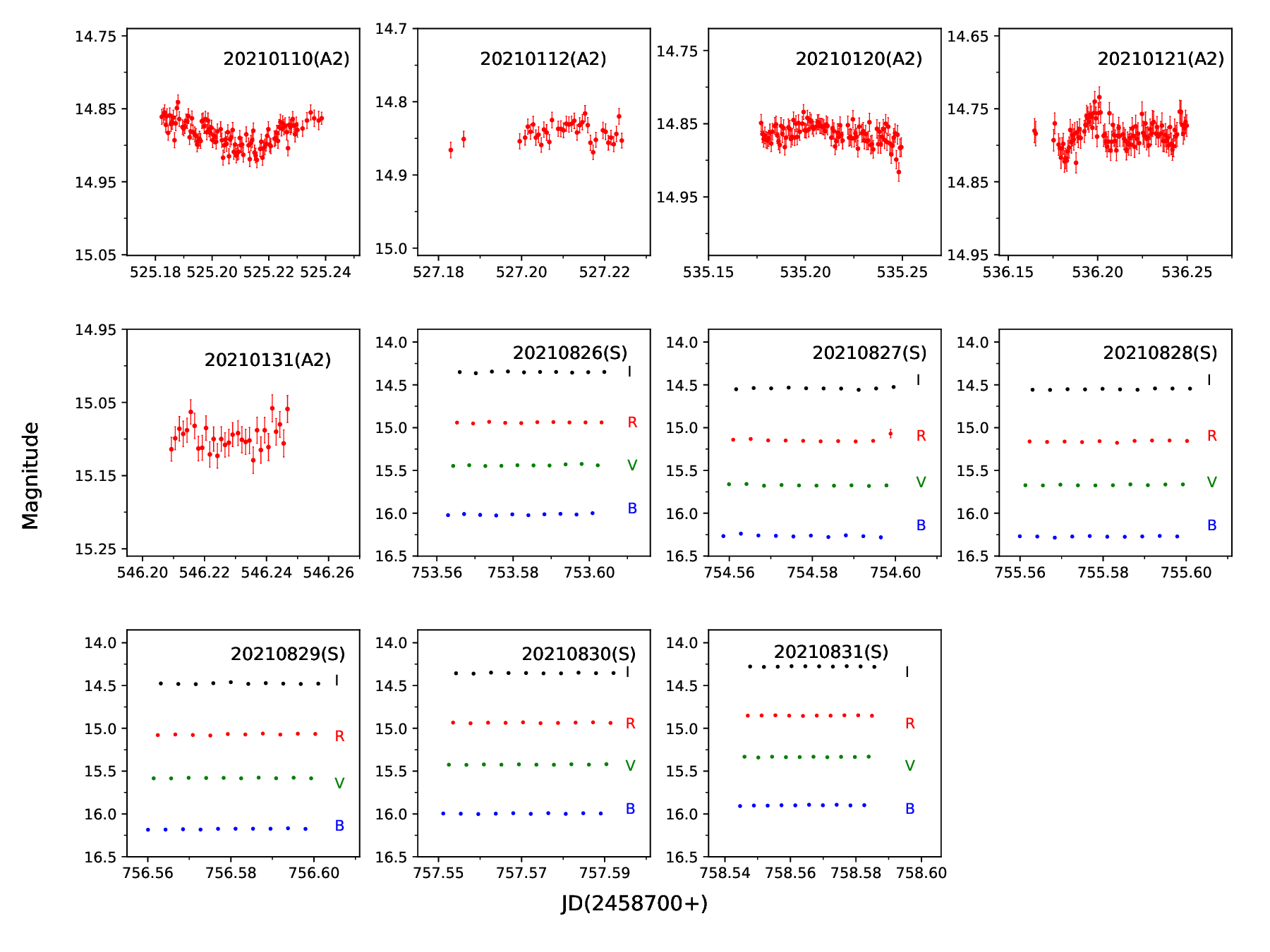}
\vspace*{-0.4in}
\caption{Continued} 
\label{fig:idv1}
\end{figure*}

\begin{table*}
\caption{Results of IDV analysis of TXS 0506$+$056.}            
\label{tab:var_res}                   
\centering 
\hskip-1.5cm  
\scalebox{0.93}{                    
\begin{tabular}{lcccccccccccc}           
\hline                		 
Observation date & Band & Duration & \multicolumn{3}{c}{{\it Power-enhanced  F-test}}  & \multicolumn{3}{c}{{\it Nested ANOVA}} &Status &   Amplitude & ACF \\
yyyy-mm-dd & &(hr)  &DoF($\nu_1$,$\nu_2$ ) & $F_{enh}$ & $F_c$  & DoF($\nu_1$,$\nu_2$ ) & $F$ & $F_c$& & $\%$& Hrs  & \\
\hline
	 	 
20191108  & R  & 1.62   & 35, 70 & 1.09 & 1.93    & 8,  27  & 5.96  & 3.26  &   NV & -- & --   \\ 
20191110  & R  & 1.21   & 23, 46 & 3.74 & 2.24    & 5,  18  & 9.88  & 4.25  &   Var & 8.69 & --   \\ 
20191116  & R  & 1.53   & 27, 54 & 1.46 & 2.11    & 6,  21  & 2.54  & 3.81  &   NV & --  & --  \\ 
20191117  & R  & 2.01   & 39, 78 & 0.37 & 1.86    & 9,  30  & 1.45  & 3.07  &   NV & --  & --  \\ 
20191221  & V  & 3.35   & 31, 62 & 0.15 & 2.01    & 7,  24  & 0.69  & 3.49  &   NV & --  & --  \\
	  & R  & 3.36   & 31, 62 & 0.75 & 2.01    & 7,  24  & 2.09  & 3.49  &   NV & --  & --  \\
	  & V-R &  & 31, 62 & 0.14 & 2.01    & 7,  24  & 0.67  & 3.49  &   NV & --  & --  \\
20191222  & R  & 4.09   & 59,118 & 1.02 & 1.66    &14,  45  & 2.09  & 2.51  &   NV & -- & --   \\
20191225  & V  & 2.93   & 31, 62 & 1.59 & 2.01	   & 7,  24  & 3.45  & 3.49  &   NV & --  & --  \\
          & R  & 2.91   & 31, 62 & 0.77 & 2.01	   & 7,  24  & 1.99  & 3.49  &   NV & --  & --  \\
          & I  & 2.96   & 31, 62 & 1.42 & 2.01	   & 7,  24  & 2.59  & 3.49  &   NV & --  & --  \\
          & V-R   &  & 31, 62 & 0.97 & 2.01	   & 7,  24  & 1.26  & 3.49  &   NV & --  & --  \\
20191226  & R  & 5.97   & 64,128 & 0.54 & 1.63	   &15,  48  & 1.76  & 2.44  &   NV & --  & --  \\
20191228  & V  & 3.21   & 29, 58 & 1.63 & 2.05    & 6,  21  & 2.89  & 3.81  &   NV & --  & --  \\ 
	  & R  & 3.19  & 29, 58 & 1.85 & 2.05    & 6,  21  & 3.41  & 3.81  &   NV & --  & --  \\
	  & V-R   &  & 29, 58 & 1.86 & 2.05    & 6,  21  & 1.25  & 3.81  &   NV & --  & --  \\ 
20191229  & V  & 3.20   & 28, 56 & 0.56 & 2.08    & 6,  21  & 2.43  & 3.81  &   NV & -- & --   \\ 	      
	  & R  & 3.20   & 28, 56 & 0.52 & 2.08    & 6,  21  & 0.96  & 3.81  &   NV & -- & --   \\ 
	  & V-R   &  & 28, 56 & 0.72 & 2.08    & 6,  21  & 1.71  & 3.81  &   NV & -- & --   \\ 
20191230  & V  & 2.94   & 27, 54 & 0.55 & 2.11    & 6,  21  & 2.55  & 3.81  &   NV & --  & --  \\
	  & R  & 2.93    & 27, 54 & 0.22 & 2.11    & 6,  21  & 0.94  & 3.81  &   NV & --  & --  \\
	  & V-R   &  & 27, 54 & 0.51 & 2.11    & 6,  21  & 1.51  & 3.81  &   NV & --  & --  \\ 
20201026  & R  & 1.03   & 67,134 & 0.29 & 1.61    &16,  61  & 6.36  & 2.37  &   NV & --  & --  \\ 
20201027  & R  & 2.21   &120,240 & 9.66 & 1.43    &29,  90  &10.08  & 1.93  &   Var & 7.75 & -- \\ 
20201112  & R  & 5.64   &195,390 & 1.85 & 1.32    &48, 147  & 5.01  & 1.68  &   Var &11.64 & -- \\ 
20201117  & V  & 3.01   & 53,106 & 1.19 & 1.71    &12,  39  & 2.29  & 2.68  &   NV &  --  & -- \\
          & R  & 3.03   & 53,106 & 1.52 & 1.71    &12,  39  & 3.47  & 2.68  &   NV & --  & --  \\
          & V-R   &   &53,106 & 0.82 & 1.71    &12,  39  & 1.01  & 2.68  &   NV & --  & --  \\
20201118  & R  & 2.36   &116,232 & 4.94 & 1.44    &28,  87  & 7.67  & 1.95  &   Var & 5.85 & --  \\
20201120  & V  & 3.89   & 55,110 & 1.81 & 1.68    &13,  42  & 9.06  & 2.59  &   Var & 15.43 & 3.50 \\
	  & R  & 3.84   & 55,110 & 3.08 & 1.68    &13,  42  & 2.81  & 2.59  &   Var & 12.98 & 3.06 \\
	  & V-R   &   &55,110 & 2.57 & 1.68    &13,  42  & 2.96  & 2.59  &   Var & 12.06 & -- \\
20201122  & R  & 2.52   &31,62 & 1.09 & 2.01      & 7,  24  & 1.16 & 3.49  &  NV & --  & -- \\
20201124  & R  & 2.82   &74, 148  & 1.78  & 1.58  & 17, 54 &  4.42 & 2.32   & Var  &  10.75 & -- \\
20201205  & R  & 1.04   & 59,118 & 0.97 & 1.66    &14,  45  & 0.76  & 2.51  &   NV & --  & -- \\
20201211  & V  & 3.47   & 39, 78 & 2.81 & 1.86    & 9,  30  & 3.38  & 3.07  &   Var & 7.34  & -- \\
          & R  & 3.48   & 39, 78 & 4.24 & 1.86    & 9,  30  & 3.61  & 3.07  &   Var & 5.45 & -- \\
	  & V-R   &   & 39, 78 & 2.58 & 1.86    & 9,  30  & 1.66  & 3.07  &   NV & --  & -- \\
20201212  & R  & 4.13   &320,640 & 1.83 & 1.25    &79, 240  & 1.32  & 1.51  &   NV & --  & --  \\
20201213  & R  & 1.02   & 87,174 & 0.95 & 1.52    &21,  66  & 1.44  & 2.14  &   NV & --  & --  \\
20210109  & R  & 1.03   & 65,130 & 0.75 & 1.62    &15,  48  & 1.76  & 2.44  &   NV & -- & -- \\
20210110  & R  & 1.36   &109,218 & 3.18 & 1.46    &26,  81  & 3.83  & 1.99  &   Var & 7.76  & --  \\
20210112  & R  & 1.01   & 35, 70 & 0.44 & 1.93    & 8,  27  & 2.14  & 3.26  &   NV & --  & -- \\
%20210115  & R    & 51 ,102& 0.23 & 1.73    &12,  39  &10.73  & 2.68  &   NV & --   \\ 
20210120  & R  & 1.71   & 99,198 & 0.95 & 1.48    & 24, 75  & 2.22  & 2.05  &   NV & -- & --  \\ 
20210121  & R  & 2.04   &104,208 & 0.39 & 1.47    & 25, 78  & 5.06  & 2.02  &   NV & -- & --  \\ 
20210131  & R  & 1.44   & 32, 64 & 0.32 & 1.98    &  7, 24  & 0.99  & 3.49  &   NV & -- & --  \\ 
20210826  & B  & 1.00      &  9, 18 & 1.83 & 3.59    &  1,  8  & 1.61  & 11.26 &   NV & -- & --  \\
          & V  & 1.00      &  9, 18 & 0.99 & 3.59    &  1,  8  & 2.81  & 11.26 &   NV & -- & --  \\
          & R  & 1.00      &  9, 18 & 0.61 & 3.59    &  1,  8  & 3.72  & 11.26 &   NV & -- & --  \\
          & I  & 1.00      &  9, 18 & 0.51 & 3.59    &  1,  8  & 0.42  & 11.26 &   NV & -- & --  \\
20210827  & B  & 1.00      &  9, 18 & 1.71 & 3.59    &  1,  8  & 0.94  & 11.26 &   NV & -- & --  \\
          & V  & 1.00      &  9, 18 & 1.19 & 3.59    &  1,  8  & 1.59  & 11.26 &   NV & -- & --  \\
          & R  & 1.00      &  9, 18 & 4.84 & 3.59    &  1,  8  & 0.19  & 11.26 &   NV & -- & --  \\
          & I  & 1.00      &  9, 18 & 2.52 & 3.59    &  1,  8  & 0.51  & 11.26 &   NV & -- & --  \\
20210828  & B  & 1.00      &  9, 18 & 0.29 & 3.59    &  1,  8  & 0.24  & 11.26 &   NV & -- & --  \\
          & V  & 1.00      &  9, 18 & 0.41 & 3.59    &  1,  8  & 9.58  & 11.26 &   NV & -- & --  \\
          & R  & 1.00      &  9, 18 & 0.56 & 3.59    &  1,  8  & 0.27  & 11.26 &   NV & -- & --  \\
          & I  & 1.00      &  9, 18 & 0.92 & 3.59    &  1,  8  & 0.59  & 11.26 &   NV & -- & --  \\
20210829  & B  & 1.00      &  9, 18 & 0.83 & 3.59    &  1,  8  & 7.47  & 11.26 &   NV & -- & --  \\
          & V  & 1.00      &  9, 18 & 0.16 & 3.59    &  1,  8  & 0.19  & 11.26 &   NV & -- & --  \\
          & R  & 1.00      &  9, 18 & 1.68 & 3.59    &  1,  8  & 5.04  & 11.26 &   NV & -- & --  \\
          & I  & 1.00      &  9, 18 & 0.38 & 3.59    &  1,  8  & 0.92  & 11.26 &   NV & -- & --  \\
20210830  & B  & 1.00      &  9, 18 & 0.21 & 3.59    &  1,  8  & 1.01  & 11.26 &   NV & -- & --  \\
          & V  & 1.00      &  9, 18 & 0.22 & 3.59    &  1,  8  & 0.87  & 11.26 &   NV & -- & --  \\
          & R  & 1.00      &  9, 18 & 1.81 & 3.59    &  1,  8  & 0.61  & 11.26 &   NV & -- & --  \\
          & I  & 1.00      &  9, 18 & 0.37 & 3.59    &  1,  8  & 0.33  & 11.26 &   NV & -- & --  \\
20210831  & B  & 1.00      &  9, 18 & 9.19 & 3.59    &  1,  8  & 1.21  & 11.26 &   NV & -- & --  \\
          & V  & 1.00      &  9, 18 & 0.94 & 3.59    &  1,  8  & 0.63  & 11.26 &   NV & -- & --  \\
          & R  & 1.00      &  9, 18 & 0.14 & 3.59    &  1,  8  & 0.88  & 11.26 &   NV & -- & --  \\
          & I  & 1.00      &  9, 18 & 0.34 & 3.59    &  1,  8  & 1.61  & 11.26 &   NV & -- & --  \\
\hline                          
\end{tabular}
}
\end{table*}

\section{ANALYSIS TECHNIQUES} \label{sec:AnaTech}

\noindent
In this section we briefly explain various analysis techniques we have used to analyse these optical data of the blazar TXS 0506+056 on diverse timescales. To obtain the blazar's intraday variability, we have examined two relatively recently developed statistical analysis techniques: the power enhanced $F$-test and the nested analysis of variance (ANOVA) test \citep{2014AJ....148...93D,2015AJ....150...44D}. These tests are usually more reliable and powerful than other statistical tests such as the standard $C$-test \citep{1999A&AS..135..477R}, $F$-test \citep{2010AJ....139.1269D}, $\chi^{2}$-test \citep{2012MNRAS.420.3147G}, ANOVA test \citep{1998ApJ...501...69D} because these involve several comparison stars \citep[but see also][]{2017MNRAS.467..340Z,2020MNRAS.498.3013Z}. 

\subsection{Power-enhanced {\it F}-test}
\label{sec:f_test}
\noindent
To explore IDV, we used the power-enhanced $F$-test following the approach of \citet{2014AJ....148...93D} and \citet{2015AJ....150...44D}. In recent studies, this test has been used for finding microvariations in blazars \citep[e.g.,][and references therein]{2015MNRAS.452.4263G,2016MNRAS.460.3950P,2017MNRAS.466.2679K,2020ApJ...890...72P,2023MNRAS.519.2796D}. In this test we compare the variance of the source light curve (LC) to the combined variance of all comparably bright comparison stars. In this work, we have three comparison field stars (A, C, and D) \citep{2021BlgAJ..34...79B} from which star C is considered as the reference star, and the remaining two field stars as the comparison stars. For details about our implementation of the power-enhanced $F$-test see \citet{2023MNRAS.519.2796D}.

\subsection{Nested {\it ANOVA}}
\label{sec:anova}
The nested ANOVA test is an updated ANOVA test that uses several stars as reference stars to generate different \blue{DLCs} of the blazar. In contrast to power-enhanced $F$-test, no comparison star is needed in the nested ANOVA test, as all the comparison stars are used as reference stars, so the number of stars in the analysis has increased by one \citep{2015AJ....150...44D,2019ApJ...871..192P}. The ANOVA test compares the means of dispersion between the groups of observations. In our case, we have used three reference stars (A, C, and D) \citep{2021BlgAJ..34...79B} to generate DLCs of the blazar. For details about the nested ANOVA test see \citet{2023MNRAS.519.2796D}. \\
\\
The results of both the statistical tests are given in Table \ref{tab:var_res}, where an LC is conservatively labeled as variable (Var) for IDV only if both the tests found significant variations in it, otherwise it is labeled as non variable (NV), though of course we cannot exclude the presence of weak intrinsic variability even in some of those cases.

\subsection{Intraday variability amplitude}
For each of the variable LCs, we calculated the flux variability amplitude (Amp) in percentage, using the standard equation given by \citet{1996A&A...305...42H}:
\begin{equation}
\label{sec:Intra}
Amp = 100\times \sqrt{(A_\mathrm{max}-A_\mathrm{min})^2 - 2 \sigma^2}
\end{equation}
Here $A_\mathrm{max}$ and $A_\mathrm{min}$ are the maximum and minimum magnitudes, respectively, in the calibrated LCs of the blazar, while $\sigma$ is the mean error. The amplitude of variability is also mentioned in the last column of Table \ref{tab:var_res} for the variable LCs.

\subsection{Discrete and auto correlation functions} \label{sec:DCF}
\noindent
The discrete correlation function (DCF) analysis is used to find possible time-lags and cross-correlations between LCs of different energy bands. This technique was introduced by \citet{1988ApJ...333..646E} and later modified by \citet{1992ApJ...386..473H} to produce better error estimates. In general, astronomical LCs are unevenly binned, and, for such LCs this technique is very useful as it can be used for unevenly sampled data. Details about the computation of the DCF we employ here are provided in \citet{2017ApJ...841..123P} and \citet{2023MNRAS.519.2796D}. In computing the DCF, one compares time series from different bands; however, when we correlate a data series with itself, we obtain the auto correlation function (ACF), which consistently exhibits a peak at $t = 0$. The presence of this prominent peak in a DCF serves as an indicator of the absence of any time lag in the data. The presence of any additional strong ACF peak indicates the presence and value of a variability timescale in the time series data \citep{2011MNRAS.413.2157R,2015A&A...582A.103G}. 

\subsection{Duty Cycle}
\noindent
The duty cycle (DC)  provides a direct estimation of the fraction of time for which a source has shown variability. We have estimated the DC of the blazar TXS 0506$+$056 by using the standard approach \citep{1999A&AS..135..477R}. For DC calculations, we considered only those LCs which were continuously monitored for $\geq$ 1 hour. For details about DC estimation, see \citet{2023MNRAS.519.2796D}.

\begin{figure*}
\centering
\includegraphics[scale=1.0]{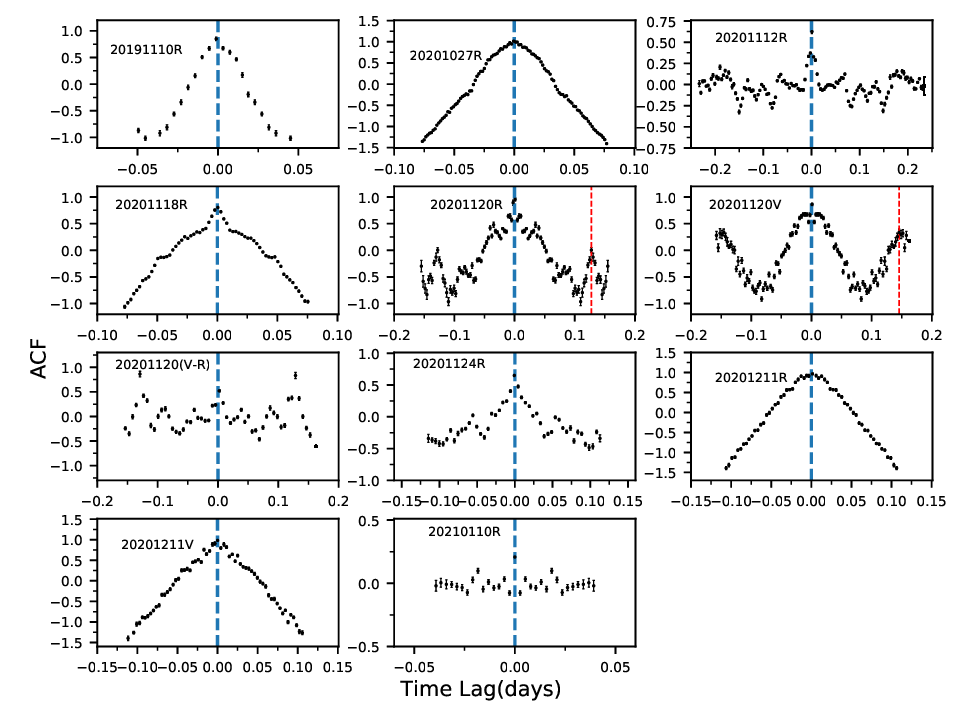}
\vspace*{-0.2in}
\caption{ACF plots of TXS 0506$+$056 light curves (including one colour index). The blue dashed lines indicate 0 lags and the red dashed lines illustrate intraday timescales.} 
\label{fig:DCF_ACF}
\end{figure*}

\section{Results}
\subsection{Intraday Flux and Colour Variability} 
\label{sec:flux}
We observed the TeV blazar TXS 0506$+$056 sufficiently intensely to investigate intraday flux variation for a span of 35 nights from November 8, 2019 to August 31, 2021. Our observations included quasi-simultaneous monitoring of the source in $V$ and $R$ bands on 7 nights, in $I$, $V$, and $R$ bands on one night, and in $B$, $V$, $R$, and $I$ bands on 6 nights. On the remaining 21 nights, we observed the source only in the $R$ band. The calibrated {\it B, V, R,} and {\it I} band IDV LCs of the blazar TXS 0506$+$056 are shown in the upper panel of each plot in Figure \ref{fig:idv1}, while the lower panels show the $V-R$ colours. IDV during some of the nights seem evident by visual inspection of the LCs. To find the presence of such rapid variability, we performed the statistical tests discussed in sections \ref{sec:f_test} and \ref{sec:anova}, and the results of the analyses are presented in Table \ref{tab:var_res}. The intra-day LCs of these 35 nights have observation duration $\geq$ 1h, and are displayed in Figure \ref{fig:idv1}. The IDV analysis results of these LCs are reported in Table \ref{tab:var_res}. However, note that the errors in $V$ band LCs are roughly twice in comparison to $R$ band, and therefore reduce the likelihood of detecting any small variations that might be present. No intraday flux variations were detected in either the {\it B} or {\it I} passbands, during the relatively few nights for which we collected sufficient measurements.  The amplitude of the IDV was estimated for the confirmed variable LCs, as shown in the second last column of Table \ref{tab:var_res}. On December 11, 2020, the {\it R} band exhibited the lowest detected variability amplitude, which was only 5.45\%, while the largest (15.43\%) variation was observed in the $V$ band on November 12, 2020. Typically, the blazar variability amplitude is larger at higher frequencies, as was seen on the one of the nights in which both the {\it R} and {\it V} band  showed variability. However, on some occasions the variability amplitude of blazars at lower frequencies has been found to be comparable to, or even larger than, that at higher frequencies \citep[e.g.,][]{2000ApJ...537..638G,2015MNRAS.452.4263G}. \\
\\
To estimate the IDV timescale of the LCs which have shown genuine variation and are listed in Table \ref{tab:var_res}, we used ACF analysis and plotted the results in Fig. \ref{fig:DCF_ACF}. IDV timescales are estimated for November 20, 2020 in $V$ and $R$ bands and listed in the last column of Table \ref{tab:var_res}. From Fig. \ref{fig:DCF_ACF}, it is seen that for the rest of the variable IDV LCs, either the timescale is longer than the data length, or ACF plots are too noisy to argue for the presence of a timescale. \\
\\
In our case there are 35 nights during which we could search for IDV (total duration is 69.75 hrs), with observations lasting between  1 $-$ 6 hrs, and we consider only those observing nights that have at least 10 data points. IDV flux variability plots are presented in Figure 1, and the analysis results are reported in Table~3. Using Eqn.\ (2) of \citet{2023MNRAS.519.2796D}, we estimated the DC values of different optical bands. We performed enhanced $F$-test and nested ANOVA tests and found 8 variable nights out of 35 in $R$-band, 2 in $V$-band from 14 nights, and so the DCs are 22.8\% and 14.3\%,  respectively. No genuine IDV was found in the $I$-band or $B$-band. We found colour variations in ($V-R$) in only 1 out of 14 observing nights and, unsurprisingly, we found no variation in $B-I$ colour. Extensive searches for optical IDV in several other TeV emitting blazars have been carried out, and it was found that either TeV blazars are not variable on IDV timescales, or if variable, the DC is in general less than 20\% \citep{2012AJ....143...23G,2012MNRAS.420.3147G,2012MNRAS.425.3002G,2016MNRAS.458.1127G,2019ApJ...871..192P,2020ApJ...890...72P,2020MNRAS.496.1430P,2023MNRAS.519.2796D}. The IDV results of the present study are in line with those earlier findings. \\
\\
We studied colour variations of the TeV blazar TXS 0506$+$056 on IDV timescales  with respect both to time and to $V$-band magnitude (colour magnitude variation). To do so,  we used the 14 nights of data on which quasi-simultaneous observations were carried out: $V$ and $R$-bands in 7 nights; $V$, $R$, $I$ bands in 1 night; and $B$, $V$, $R$, $I$ bands in 6 nights. We calculated the $V-R$ colour indices (CIs)  for each pair of $V$ and $R$ magnitudes and  plotted these $V-R$ CIs with respect to time in the bottom panel of Figure \ref{fig:idv1}. We performed the enhanced $F$-test and nested ANOVA test on each night's data set and we found $V-R$ colour variation on only 1 night, November 20, 2020. 

\begin{figure}
\centering
\includegraphics[scale=0.21]{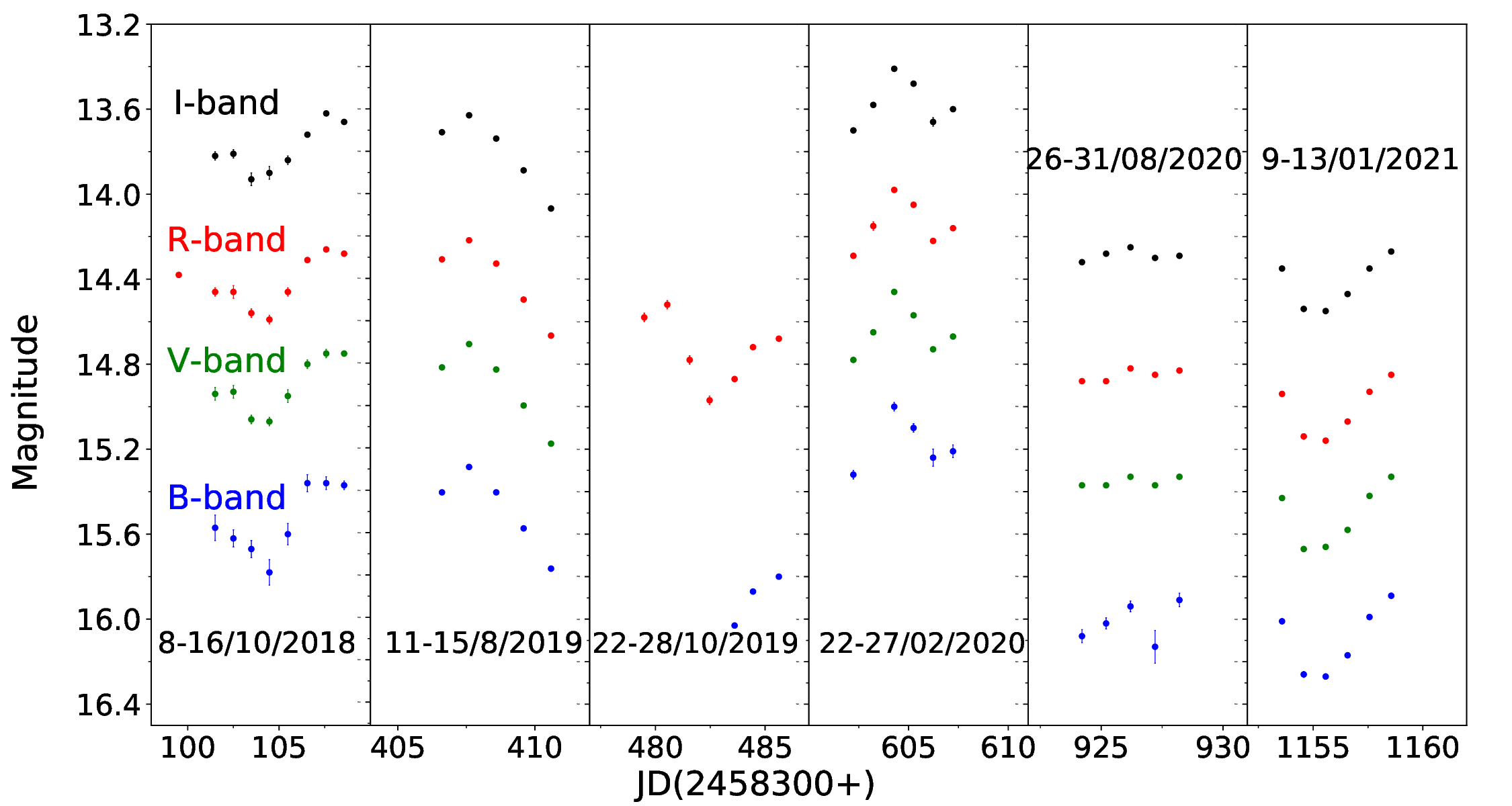}
\caption{The STV LC of TXS 0506$+$056 in $I$, $R$, $V$, and $B$ bands.} 
\label{fig:STV}
\end{figure}

\begin{table}
\caption{Short term LCs values}            
\label{tab:ST_values}                   
\centering    
\scalebox{0.8}{  
\hskip-0.5cm              
\begin{tabular}{lccccc}           
\hline                		 
ST & Duration &  Brightest  & Faintest &  Mean     & Amplitude  \\
   & yyyymmdd &  Magnitude  & Magnitude& Magnitude & Variation($\%$) \\		 

\hline 
ST1	&20181008-20181016 	& 14.26$\pm$0.01   & 14.59$\pm$0.03  & 14.42$\pm$0.03  &  32.87	 \\ 
ST2	&20190811-20190815	& 14.22$\pm$0.01   & 14.67$\pm$0.02  & 14.41$\pm$0.02  &  44.94	 \\          
ST3	&20191022-20191028 	& 14.52$\pm$0.02   & 14.97$\pm$0.02  & 14.73$\pm$0.02  &  44.91	 \\
ST4	&20200222-20200227 	& 13.98$\pm$0.01   & 14.29$\pm$0.03  & 14.14$\pm$0.03  &  30.87	 \\
ST5	&20210109-20210113 	& 14.82$\pm$0.02   & 14.88$\pm$0.01  & 14.85$\pm$0.02  &  ~5.61	 \\
ST6	&20210827-20210901 	& 14.85$\pm$0.01   & 15.16$\pm$0.03  & 15.02$\pm$0.03  &  30.87	 \\
\hline
\end{tabular}
}
\end{table}

\subsection{Short and Long Term Variability}
\subsubsection{Short Term Flux Variability}
\label{sec:STFV}
We divided the long term LC of TXS 0506$+$056  into 6 Short term LCs, ST1, 
 ST2, ST3, ST4, ST5, and ST6, as plotted in Fig. \ref{fig:STV} and the variability results are listed in Table \ref{tab:ST_values}.  The lengths of these short term LCs, along with their brightest and faintest magnitudes, mean magnitude, and amplitude of flux variations the in $R$-band are also given there. 

\subsubsection{Long Term Flux Variability}
\label{sec:LTFV}
\begin{figure*}
\centering
\includegraphics[scale=1]{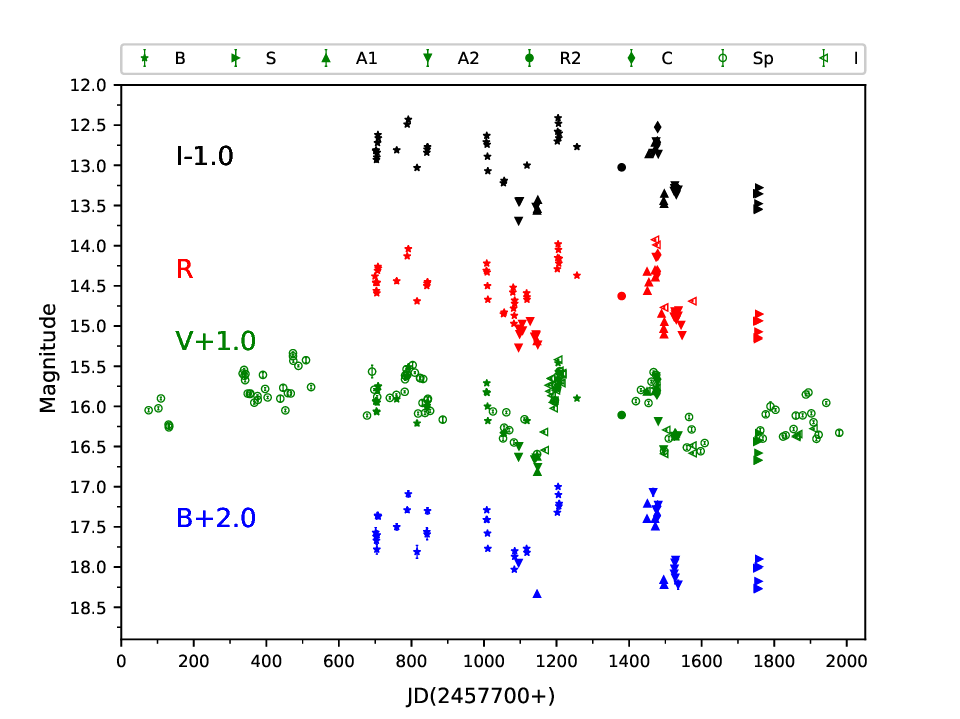}
\caption{LTV optical ($BVRI$) light curves of TXS 0506$+$056.}
\label{fig:stv_ltv}
\end{figure*}

Figure \ref{fig:stv_ltv} illustrates the long-term (LT) LCs of TXS 0506+056 in the {\it B, V, R,} and {\it I} bands over the entire monitoring period. The plot depicts the nightly averaged magnitudes in the respective bands as a function of time. During our monitoring period the source was detected in the brightest state of $R = 13.98$\,mag on February 24, 2020 while the faintest level detected was {\it R} = 15.16 mag on August 29, 2021. The mean magnitudes were 15.62, 15.02, 14.55 and 13.91 in {\it B, V, R}, and {\it I} bands, respectively. The presence of variability on LT timescales is clearly evident across all optical wavelengths. Using Equation \ref{sec:Intra}, we have estimated very similar variability amplitudes of 117.6\%, 137.9\%, 117.6\%, and 113.8\% in {\it B, V, R,} and {\it I} bands, respectively.

\begin{figure}
\centering
\includegraphics[scale=0.53]{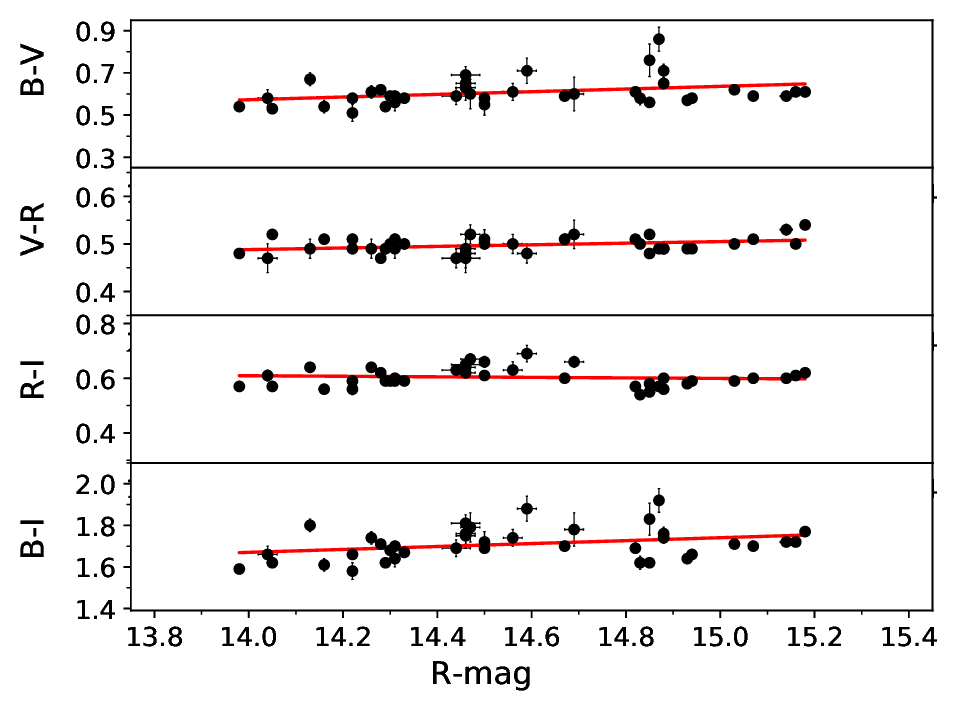}
\vspace*{-0.1in}
\caption{Optical colour-magnitude plot with respect to $R$-band of TXS 0506$+$056}
\label{fig:ST_SI1}
\end{figure}

\begin{figure}
\centering
\includegraphics[scale=0.53]{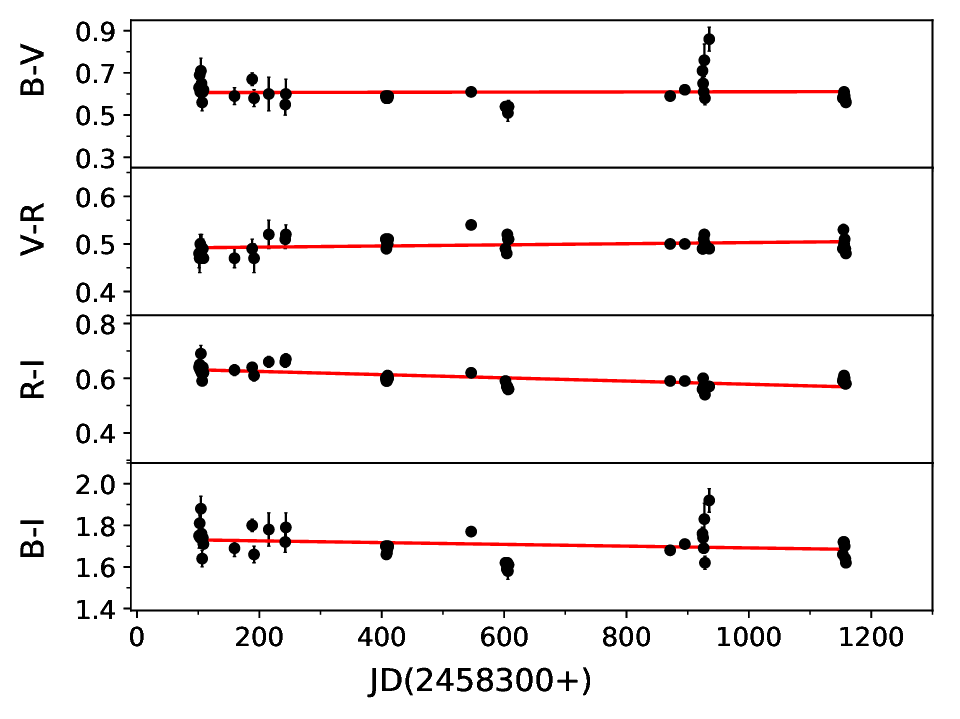}
\vspace*{-0.1in}
\caption{Optical colour variability LCs covering the total observation duration of TXS 0506$+$056}
\label{fig:ST_SI2}
\end{figure}

\begin{table}
\caption{Colour–Magnitude Dependencies and Colour–Magnitude Correlation Coefficients on LTV Timescales}
\label{tab:CI_R}                   
\centering    
\resizebox{0.45\textwidth} {!}{  
%\hskip-1.cm              
\begin{tabular}{ccccc} \hline \hline                		 
Colour Index 	& $m_1^a$        				&  $c_1^a$      	& $r_1^a$      	&  $p_1^a$  \\\hline 
$B-I$	    	&	0.083$\pm$ 0.035 	       	&0.496 	       	&0.346 	     	    &0.021			\\
$R-I$			&	0.003$\pm$ 0.019 	 		&0.544			&0.028 			    &0.856			\\
$V-R$			&	0.013$\pm$ 0.007			&0.313 			&0.238			&0.118			\\
$B-V$ 			&	0.066$\pm$ 0.033		 	&-0.367	 		&0.297			&0.051			\\
\hline                           
\end{tabular}}\\
$^am_1$ = slope and $c_1$ = intercept of CI against $R$-mag; $r_1$ = Correlation coefficient; $p_1$ = null hypothesis probability
\end{table}

\begin{table}
\caption{Colour Variation with Respect to Time on LTV Timescales}
\label{tab:CI_JD}                   
\centering    
\resizebox{0.45\textwidth} {!}{  
\begin{tabular}{ccccc} \hline \hline                		 
Colour Index 	& $m_1^a$ 						&  $c_1^a$	& $r_1^a$ 		&  $p_1^a$  \\\hline 
$B-I$			&-4.251e-05$\pm$3.238e-05 		&1.724 		&-0.199  		&0.196		\\ 
$R-I$ 			&-6.613e-05$\pm$1.418e-05		&0.641 	    &-0.584   		&3.174e-05		\\	 
$V-R$			&1.187e-05$\pm$7.137e-06	    &0.491 		&0.249 			&0.104   \\		 
$B-V$			&1.172e-05$\pm$3.086e-05    	&0.593	 	&0.058   		&0.706		 \\ \hline
\end{tabular}}\\
$^am_1$ = slope and $c_1$ = intercept of CI against time; $r_1$ = Correlation coefficient; $p_1$ = null hypothesis probability
\end{table}

\subsection{Spectral Variability}
\label{sec:SV}
The colour-magnitude (CM) relationship serves as a valuable tool for investigating different variability scenarios and gaining a deeper understanding of the origin of blazar emission. In this study, we conducted a search to identify any potential relationship between the source's colour indices (CIs) with brightness in the $R$-band and with respect to time. We fitted the plots of the optical ($B-V$), ($B-I$), ($V-R$) and ($R-I$) CIs with respect to both $R$-band magnitude and time using  straight lines of the form Y = mX + C as shown in Figs.~\ref{fig:ST_SI1} and \ref{fig:ST_SI2}, respectively \citep{2020ApJ...890...72P}. The values of the parameters related to the colour-time and colour-magnitude plots are respectively provided in the accompanying Tables \ref{tab:CI_R} and \ref{tab:CI_JD}. In our analysis, we observed modestly significant variations ($p < 0.51$) in the $B-V$ and $B-I$ colours with respect to the $R$ magnitude. On the other hand, the CIs involving the $R$-band exhibited very weak trends in the same directions.  A positive slope defines a positive correlation between CIs and blazar $R$ magnitude, meaning that the source follows a bluer-when-brighter (BWB) trend, while a negative slope defines redder-when-brighter (RWB) trend \citep[e.g.,][and references therein]{2017MNRAS.465.4423G}. We find a negative correlation of the ($R-I$) CI with time while the $B-I$ colour shows a very weak (about 1.5$\sigma$) negative slope.  The $B-V$ CI is essentially constant while the $V-R$ one shows a very weak positive trend with time.

\begin{figure}
\centering
\includegraphics[scale=0.5]{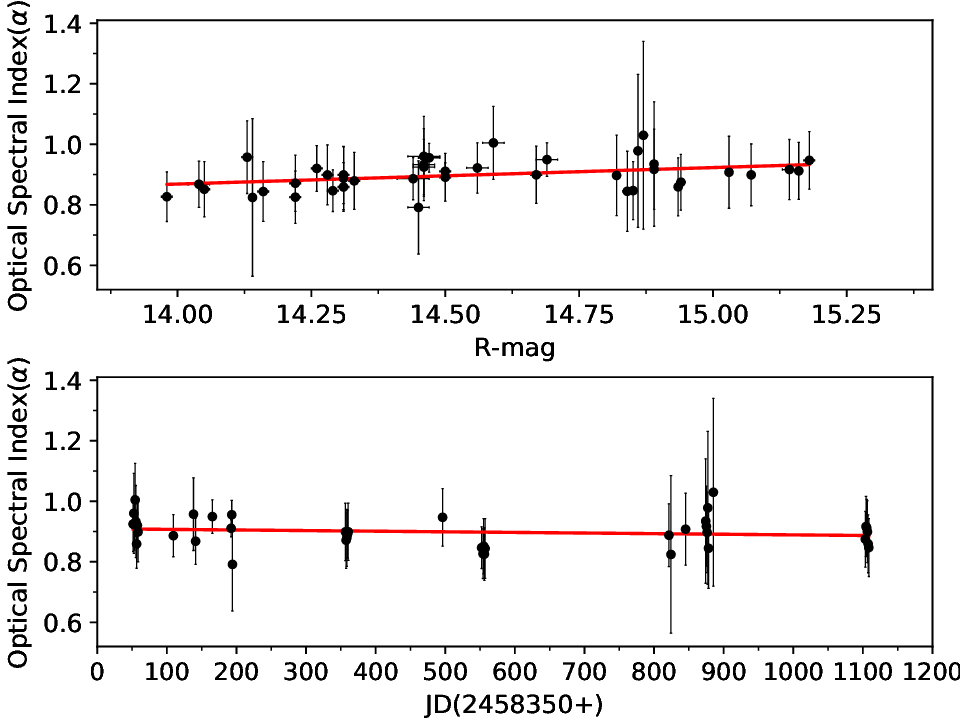}
\caption{Variation of optical spectral index ($\alpha$) of TXS 0506$+$056 with respect to $R$-magnitude (top) and JD (bottom).} 
\label{fig:SED_variation}
\end{figure}

\begin{table}
\caption {Variation of optical spectral index ($\alpha$), with respect to $R$-magnitude and LTV timescale}
\label{tab:Alpha_T_JD}                   
\centering    
\resizebox{0.45\textwidth} {!}{  
\begin{tabular}{ccccc} \hline \hline                		 
Parameters              & $m$        &  $c$         & $r$           &$p$        \\\hline 
$\alpha$ vs Time   	&-2.26e-05$\pm$1.08e-05 	 &0.909 		&-0.172   		&0.264    \\ 
$\alpha$ vs R$_{mag}$ &0.053$\pm$0.022 	 &0.123 	    &0.362	    	&0.016    \\\hline
\end{tabular}}\\
$m$ = slope and $c$ = intercept of $\alpha$ against R-mag and JD; $r$ = Correlation coefficient; $p$ = null hypothesis probability
\end{table}

\subsection{Spectral Energy Distribution (SED)} 
\label{sec:spec}
Throughout our observation period, we extracted the optical ($BVRI$) SEDs of the blazar on 44 nights. These SEDs were derived from quasi-simultaneous observations conducted across all four wavebands. For this, we first dereddened the calibrated $B$, $V$, $R$, and $I$ magnitudes by subtracting the Galactic extinction, with $A_{\lambda}$ having the following values: $A_B$ = 0.392 mag, $A_V$ = 0.297 mag, $A_R$ = 0.235 mag, and $A_I$ = 0.163 mag. The values of $A_{\lambda}$ were taken from the NASA Extragalactic Database (NED)\footnote{\url{https://ned.ipac.caltech.edu/}}. After applying the necessary dereddening and calibration procedures, the magnitudes in each band were converted into corresponding flux densities $F_{\nu}$ that had been corrected for extinction. The optical SEDs of TXS 0506$+$056, in log($\nu$) versus log($F_{\nu}$) representation, are plotted in Figure \ref{fig:sed1}. We measured the source's faintest and brightest fluxes on  August 29, 2021 and February 24, 2020 respectively. \\
\\
To determine the optical spectral indices, we fitted each SED with a first-order polynomial model of the form log($F_{\nu}$) = $-\alpha$ log($\nu$) + C. The obtained fitting results are presented in Table \ref{tab:SED_T}. The values of the spectral indices (${\alpha}$) range from 0.791 $\pm$ 0.154 to 1.029 $\pm$ 0.194 and their mean was 0.897 $\pm$ 0.171. This mean value of the spectral index closely matches  previously reported results for TXS 0506$+$056 \citep{2021ApJ...908..113H,2018ApJ...854L..32P}. \\ 
\\
In Figure \ref{fig:SED_variation}, the top and bottom panels display the spectral indices of TXS 0506$+$056 with respect to time and $R$-band magnitude, respectively. We fitted each panel in Figure \ref{fig:SED_variation} with a first-order polynomial to investigate any variations in the spectral index. The corresponding fitting parameter values are provided in Table \ref{tab:Alpha_T_JD}. The optical spectral index demonstrates a hint of a decreasing trend with time and apparently exhibits a weak positive correlation with the $R$-band magnitude. However, given the slope of the variation (0.053) and the relatively large error on it (0.022) we cannot have confidence in the reality of this trend. 

\begin{figure}
\centering
\includegraphics[scale=0.58]{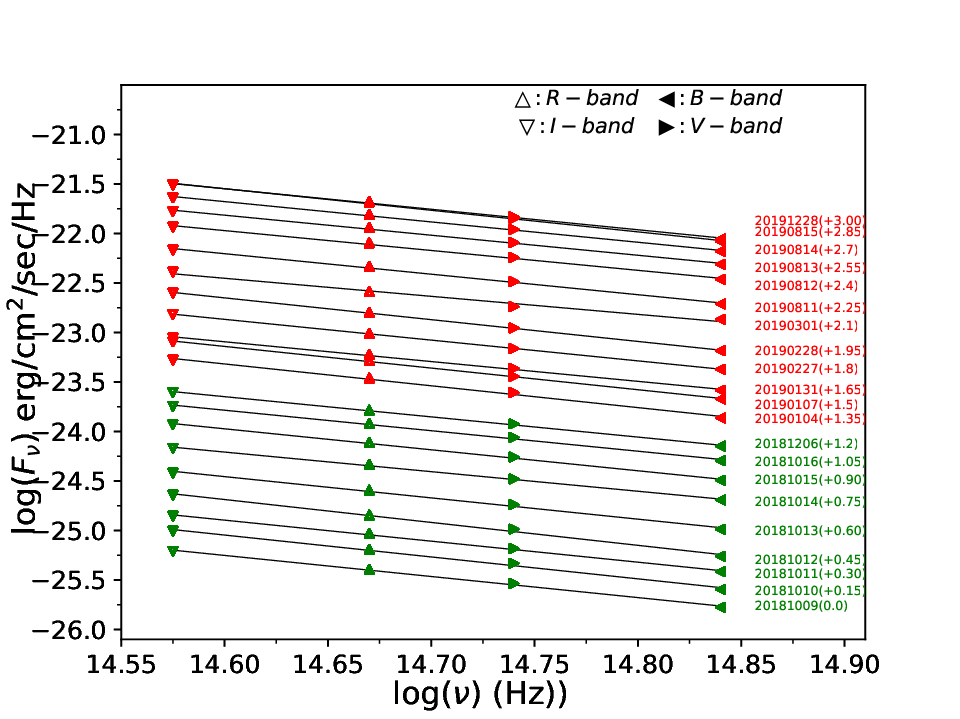}
\vspace*{-0.3in}
\caption{\label{4_a}The SED of TXS 0506$+$056 in $I$, $R$, $V$, and $B$ bands.} \label{fig:sed1}
\end{figure}

\begin{figure}
\ContinuedFloat
\centering
\includegraphics[scale=0.58]{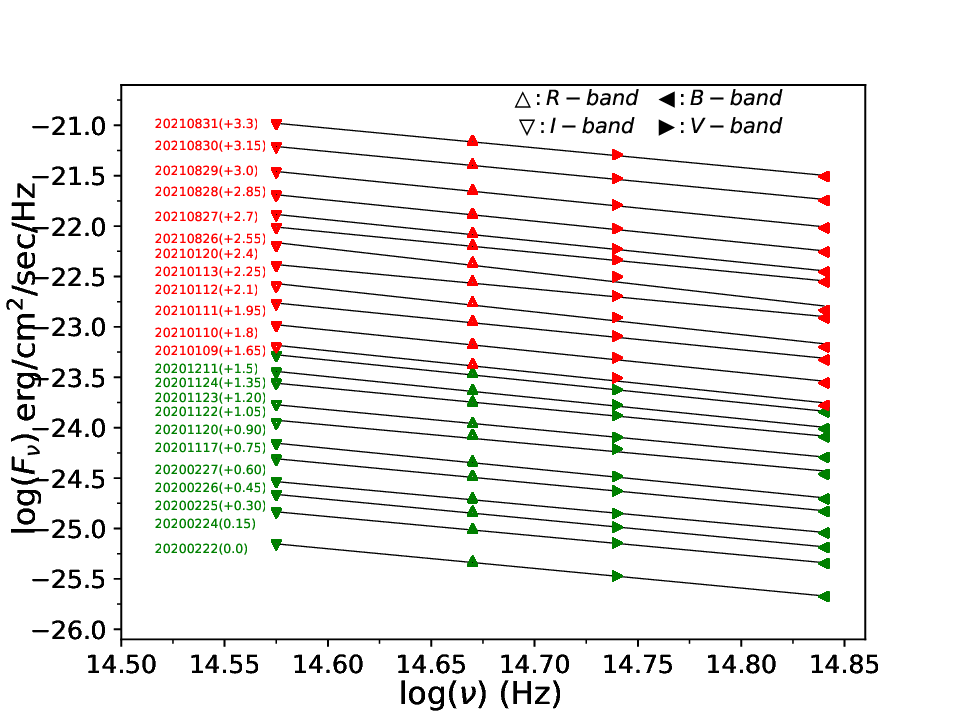}
\vspace*{-0.3in}
\caption{\label{4_b} Continued} 
\end{figure}

\begin{table*}
\caption{Straight-line Fits to Optical SEDs of TeV Blazar TXS 0506$+$056}    \scalebox{1.0}{  
\label{tab:SED_T}                   
\centering    
\hskip-1.cm              
\begin{tabular}{lccccccccc}           
\hline                		 
Observation date & $m_1^a$ &  $c_1^a$ & $r_1^a$ &  $p_1^a$  & Observation date & $m_1^a$ &  $c_1^a$ & $r_1^a$ &  $p_1^a$  \\
yyyy-mm-dd   &         &          &         &               & yyyy-mm-dd   &         &          &         &               \\		 
\hline 
2018-10-09	&0.925$\pm$0.091 	& 2.539   & -0.998  & 0.002	&	2020-02-24 	&0.827$\pm$0.082	& 1.179   & -0.998  & 0.002 \\ 
2018-10-10	&0.961$\pm$0.132	& 3.145   & -0.996  & 0.004	&	2020-02-25 	&0.852$\pm$0.091	& 1.616   & -0.998  & 0.002 \\   
2018-10-11	&0.922$\pm$0.083 	& 2.651   & -0.998  & 0.002	&	2020-02-26 	&0.825$\pm$0.086	& 1.289   & -0.998  & 0.002 \\
2018-10-12	&1.005$\pm$0.121 	& 3.949   & -0.997  & 0.003	&	2020-02-27 	&0.844$\pm$0.098 	& 1.659   & -0.997  & 0.003 \\
2018-10-13	&0.933$\pm$0.119 	& 3.009   & -0.997  & 0.003 &	2020-11-17 	&0.887$\pm$0.103 	& 2.365   & -0.997  & 0.003 \\
2018-10-14	&0.859$\pm$0.081 	& 2.031   & -0.998  & 0.002	&	2020-11-20 	&0.824$\pm$0.261 	& 1.539   & -0.981  & 0.018 \\
2018-10-15 	&0.921$\pm$0.076 	& 3.027   & -0.999  & 0.001	&	2020-11-22 	&0.849$\pm$0.119 	& 1.969   & -0.997  & 0.003 \\
2018-10-16 	&0.898$\pm$0.098 	& 2.795   & -0.998  & 0.002	&	2020-11-23 	&0.859$\pm$0.205 	& 2.213   & -0.991  & 0.009 \\
2018-12-06 	&0.886$\pm$0.069 	& 2.672   & -0.999  & 0.001	&	2020-11-24 	&0.917$\pm$0.132 	& 3.174   & -0.996  & 0.004 \\
2019-01-04	&0.957$\pm$0.119 	& 3.853   & -0.997  & 0.003	&	2020-12-11 	&0.907$\pm$0.119 	& 2.964   & -0.996  & 0.004 \\
2019-01-07 	&0.868$\pm$0.076 	& 2.646   & -0.999  & 0.001	&	2021-01-09 	&0.935$\pm$0.205 	& 3.468   & -0.997  & 0.012 \\
2019-01-31	&0.949$\pm$0.055 	& 3.814   & -0.999  & 0.001	&	2021-01-10 	&0.917$\pm$0.132 	& 3.305   & -0.996  & 0.004 \\
2019-02-27 	&0.911$\pm$0.028 	& 3.368   & -0.999  & 0.001	&	2021-01-11 	&0.897$\pm$0.133 	& 3.106   & -0.995  & 0.017 \\
2019-02-28 	&0.955$\pm$0.048 	& 4.116   & -0.999  & 0.001	&	2021-01-12 	&0.978$\pm$0.152 	& 4.371   & -0.998  & 0.002 \\
2019-03-01 	&0.791$\pm$0.154 	& 1.807   & -0.993  & 0.007	&	2021-01-13 	&0.845$\pm$0.132	& 2.507   & -0.995  & 0.002 \\
2019-08-11 	&0.898$\pm$0.095 	& 3.486   & -0.998  & 0.002	&	2021-01-20 	&1.029$\pm$0.194 	& 5.298   & -0.991  & 0.002 \\
2019-08-12 	&0.871$\pm$0.093 	& 3.182   & -0.998  & 0.002	&	2021-08-26 	&0.875$\pm$0.097 	& 3.107   & -0.998  & 0.002 \\
2019-08-13 	&0.879$\pm$0.094 	& 3.363   & -0.998  & 0.002	&	2021-08-27 	&0.917$\pm$0.099 	& 3.771   & -0.998  & 0.002 \\
2019-08-14 	&0.891$\pm$0.079 	& 3.599   & -0.999  & 0.001 &	2021-08-28 	&0.912$\pm$0.095 	& 3.791   & -0.998  & 0.002 \\
2019-08-15 	&0.899$\pm$0.095 	& 2.739   & -0.998  & 0.002 &	2021-08-29 	&0.899$\pm$0.102 	& 3.699   & -0.998  & 0.002 \\
2019-12-28 	&0.947$\pm$0.095	& 4.469   & -0.998  & 0.002 &   2021-08-30 	&0.859$\pm$0.096 	& 3.231   & -0.998  & 0.002 \\
2020-02-22 	&0.847$\pm$0.069	& 3.242   & -0.998  & 0.001 &   2021-08-31 	&0.847$\pm$0.095 	& 3.148   & -0.998  & 0.002 \\
\hline                           
\end{tabular}
} \\
$^am_1$ = slope and $c_1$ = intercept of log($F_{\nu}$) and log(${\nu}$); $r_1$ = Correlation coefficient; $p_1$ = null hypothesis probability\\
\end{table*}

\section{Discussion and Conclusions}
The thermal radiation in blazars emitted by the accretion disc is typically overwhelmed by the Doppler-boosted nonthermal radiation originating from the relativistic jet.  As a result, any observed variability is more likely to be explained by models based on the relativistic jet \citep[e.g.,][and references therein]{2015MNRAS.450..541A}. During periods when a blazar exhibits a very low flux state, the observed variability could potentially be attributed to hotspots or instabilities occurring on the accretion disc \citep[e.g.,][]{1993ApJ...411..602C,1993ApJ...406..420M}. IDV/STV observed in the optical bands can be attributed to the presence of turbulence near a  shock in the jet, as well as other irregularities within the jet flow resulting from variations in the outflow parameters \citep[e.g.,][and references therein]{2014ApJ...780...87M,2015JApA...36..255C}.\\
\\
The micro-level flux variations observed in blazars LCs on IDV timescales may be attributed to the turbulent plasma flowing at relativistic speeds within the jet \citep[e.g.,][]{2014ApJ...780...87M,2016ApJ...820...12P} or to mini-jets within the jets \citep{2009MNRAS.395L..29G}. Different optical IDV behaviours have been observed in two subclasses of blazars, namely LBLs and HBLs. HBLs display relatively less variability in optical bands on IDV time-scales compared to their amplitude of variability in X-rays and $\gamma$-rays \citep[e.g.,][and references therein]{1996A&A...305...42H,2011MNRAS.416..101G,2012MNRAS.420.3147G,2016MNRAS.462.1508G}. The presence of strong magnetic fields within the relativistic jet may be responsible for the different microvariability behaviours observed in the optical bands of both LBLs and  HBLs \citep{1999A&AS..135..477R}. The idea is that stronger magnetic fields in HBLs can potentially interrupt the formation of small fluctuations caused by Kelvin-Helmholtz instabilities within relativistic jets. These fluctuations typically interact with shocks in the jets, leading to IDV. However, the generation of very rapid variability can be disrupted if the strength of the magnetic field exceeds the critical value $B_{c}$ \citep{2005ChJAS...5..110R}, 
\begin{equation}
B_c = \big[4\pi n_e m_e c^2(\gamma^2 - 1)\big]^{1/2} \gamma^{-1}
\end{equation}
where $n_e$ is the local electron density, $m_e$ is the rest mass of the electron, and $\gamma$ is the bulk Lorentz factor of the flow. The bulk Lorentz factor $\gamma$, and the Doppler factor $\delta$ are given by $\delta = [\gamma (1 - \beta cos \ \theta]^{-1}$, where $\beta$c and $\theta$ are the velocity of emitting plasma and jet viewing angle, respectively. For TXS 0506+056, $\theta, \delta, \rm{and} \ \gamma$ are is found to be 8$^{\circ}$ -- 20$^{\circ}$, 2 -- 9, and $\sim$ 5, respectively  \citep{2019MNRAS.483L..42K,2020ApJ...896...63L,2022MNRAS.509.1646S}. The local electron density $n_e$ can be estimated using broadband multi-wavelength SED of blazars. However, for TXS 0506+056 a broadband multi-wavelength SED is not available, so we took the average value of $n_e$ for other possible potential neutrino loud TeV blazars (e.g., Mrk 421, Mrk 501, 1ES 1426+428, PKS 2155-304, 1ES 2344+514, etc.) \citep{2002PhRvD..66l3003N}. The electron density $n_e$ of these blazars are found to be in the range of 0.3 -- 1.45 electrons cm$^{-3}$ \citep{2011ApJ...727..129A,2011ApJ...736..131A,2012AIPC.1505..635C,2019ICRC...36..768P,2020MNRAS.496.3912M}. Considering these parameter values, we get B $\approx$ 0.1 G. As TXS 0506+056 is a TeV blazar, it is expected to have $B > B_{c}$ that, within this scenario, would inhibit the development of the small scale structures  that could yield IDV in the optical LCs. \\
\\
Changes in colour or spectral index can aid in understanding of the emission mechanisms in blazars. As mentioned earlier, three distinct types of behaviour can be observed in the colour-magnitude diagram (CMD): redder-when-brighter (RWB), bluer-when-brighter (BWB), and achromatic. The BWB trend is commonly observed in BL Lacs, whereas the RWB trend is typically followed by FSRQs \citep[e.g.,][]{2012AJ....143...23G,2015MNRAS.452.4263G}. Synchrotron models dominated by  one component can explain the BWB behaviour if the energy distribution of injected fresh electrons, which cause an increase in flux, is harder \citep{1998A&A...333..452K}. Another possible explanation for BWB behaviour involves precession of the jet, causing variations in the Doppler factor that affect the `convex' spectrum  \citep{2004A&A...421..103V}. During our observations, we consistently observed that TXS 0506$+$056 followed the BWB trend. This suggests that the increasing flux can be attributed to jet synchrotron emission, very possibly indicating an enhancement in particle acceleration efficiency \citep{2015MNRAS.451.3882A}. This BWB trend is seen in Figure \ref{fig:SED_variation} which illustrates spectral steepening as the magnitude increases. This trend can also be explained by a two component emission picture: one is a stable component ($\alpha_{\rm{constant}}$) while the other is a variable component with a flatter slope ($\alpha_{1}$). When the variable component dominates over the stable component then chromatic behaviours are exhibited.  On shorter timescales, optical variations are predominantly influenced by strong chromatic components, while longer-term optical variations can be attributed to a mildly chromatic component \citep{2004A&A...424..497V}.\\ 
\\
The spectral and flux variations observed in TXS 0506$+$056 may offer valuable insights into the evolutionary processes occurring in radio loud AGN \citep[e.g.,][]{2002ApJ...566L..13B,2003ApJ...585L..23F}. In an accretion disc-based model, the shortest variability time-scale is connected to the time it takes for light to traverse the variable region which is directly related to the BH mass  \citep{2005MNRAS.358..774B}. The likelihood of detecting variability in blazars becomes greater as the duration of observations increases. For instance, extending the observation period from $\leq$ 3 to $\geq$ 6 hours resulted in an increase in the probability of identifying IDV in blazars from 64\% to 82\% \citep{2005A&A...440..855G}. \\
\\
In this study, we conducted an analysis of optical photometric data obtained from seven ground-based telescopes, covering the period from October 2018 to August 2021, for the TeV blazar TXS 0506$+$056. This study represents the first comprehensive investigation of the temporal and spectral behaviour of this source across a wide range of timescales within the optical domain. We examined a total of 35  $R$-band, 14 $V$-band, 7 $I$-band, and 6 $B$-band IDV LCs using two powerful and robust statistical methods, namely the nested ANOVA test and the power-enhanced $F$-test. During the period of our observations, we observed significant variability in the $R$-band on 8 nights and in the $V$-band on 2 nights. However, no IDV was detected in the $B$ and $I$ bands for which we took many fewer measurements. Additionally, we observed a variation in the ($V-R$) colour  within only one night throughout the entire observation period. The Duty Cycles (DCs) we measured for the {\it R} and {\it V} bands are 22.8\%, and 14.3\%, respectively. So, from our IDV analysis, we conclude that optical LCs of TXS 0506$+$056 are either constant or show nominal variations on IDV time-scales. The blazar TXS 0506$+$056 did not show large-amplitude variations during our monitoring period. One important caveat is that the duration of our observations ranges from 1 to 6 hours, so it is certainly possible that if more of our nightly observations had covered longer periods we would have seen more frequent IDV. In a statistical study of the optical IDV of blazars, it was found that chances of detection of IDV is 60 -- 65\% if observations are performed for less than 6h, but if the blazar is observed for more than 6h, the chance of detecting IDV is 80 -- 85\% \citep{2005A&A...440..855G}.  \\
\\
By analysing the ACFs, we were able to identify evidence of variability timescales on November 20, 2020, in both the $V$ and $R$ bands. The variability timescales were determined to be 3.50 hours in the $V$ band and 3.06 hours in the $R$ band reported in Table \ref{tab:var_res} and shown in Figure \ref{fig:DCF_ACF}. To obtain an upper limit for the size ($R$) of the emission region, we apply the simple causality constraint,
\begin{equation}
R \leq \frac{c ~t_\mathrm{var} ~\delta}{1+z} 
\end{equation}
\noindent
where $\delta$ is the Doppler factor. The values of  $\delta$ for TXS 0506+056 have been estimated to be between 2 and 9 using radio very long baseline interferometry (VLBI) / very long baseline array (VLBA) data in the frequency range 8 GHz to 43 GHz \citep{2019MNRAS.483L..42K,2020ApJ...896...63L,2022MNRAS.509.1646S}. Using that range of Doppler factor values, taking $z = 0.3365$ \citep{2018ApJ...854L..32P}, and employing the shortest variability timescale we found ($t_\mathrm{var}$ = 11.02 ks in the $R$ band), and employing Eqn.\ (3), we obtain that the size of the emission region is in the range of 4.95 $\times \rm{10}^{14}$cm -- 2.23 $\times {10}^{15}$cm. \\
\\
To determine the optical spectral index ($\alpha$), we constructed optical SEDs using quasi-simultaneous observations in the $B$, $V$, $R$, and $I$ bands at different epochs. The analysis revealed a weak positive correlations between the optical spectral index ($\alpha$) and $R$-band magnitude. Over the Long-Term Variability (STV) timescale, the weighted mean value of $\alpha$ was determined to be 0.897 $\pm$ 0.171. Since TXS 0506+056  is the best case for a blazar producing neutrino fluxes it should remain a priority target for continuing observations in all bands. 

\section*{Acknowledgements}
We thank the anonymous reviewer for very useful comments that helped us to improve the manuscript. ACG is partially supported by Chinese Academy of Sciences (CAS) President’s International Fellowship Initiative (PIFI) (grant no. 2016VMB073). This research was partially supported by the Bulgarian National Science Fund of the Ministry of Education and Science under grants KP-06-H28/3 (2018), KP-06-H38/4 (2019), KP-06-KITAJ/2 (2020) and KP-06-H68/4 (2022). The Skinakas Observatory is a collaborative project of the University of Crete, the Foundation for Research and Technology -- Hellas, and the Max-Planck-Institut f\"{u}r Extraterrestrische Physik. HG acknowledges financial support from the Department of Science and Technology (DST), Government of India, through INSPIRE faculty award IFA17-PH197 at ARIES, Nainital, India. Based on data acquired at Complejo Astron\'omico El Leoncito, operated under agreement between the Consejo Nacional de Investigaciones Cientificas y T\'ecnicas de la Rep\'ublica Argentina and the National Universities of La Plata, C\'ordoba, and San Juan.

\section*{Data Availability}
The data of this article will be shared after one year of the publication of the paper at a reasonable request to the first author.

\end{document}